\documentclass[runningheads]{llncs}
\usepackage[T1]{fontenc}
\usepackage{listings}
\usepackage{xcolor}
\usepackage{float}
\usepackage{caption}    
\usepackage{placeins}
\usepackage[htt]{hyphenat} 
\usepackage{subcaption}
\usepackage{seqsplit}
\usepackage{fancyvrb}
\usepackage[ruled,vlined,linesnumbered,longend]{algorithm2e}
\SetAlgoSkip{}
\usepackage{amsmath}
\usepackage{booktabs}

\lstdefinelanguage{Go}{
  morekeywords={package,import,func,return,if,for,range,var,const,type,struct,interface,select,case,break,continue,go,chan,map},
  sensitive=true,
  morecomment=[l]{//},
  morecomment=[s]{/*}{*/},
  morestring=[b]"
}

\usepackage{graphicx}
\usepackage{hyperref}
\begin{document}
\title{Beyond Takedown: Measuring Malicious Go Module Persistence in the Wild}
\author{Minjae Bae \and Carter Yagemann}
\authorrunning{M. Bae and C. Yagemann}
\institute{The Ohio State University, Columbus, OH, USA\\
\email{\{bae.299,yagemann.1\}@osu.edu}}

\maketitle

\begin{abstract}
We measure an automation-based supply chain campaign in the Go ecosystem. The attackers repackage legitimate Go modules under attacker-controlled owners, and embed them with obfuscated code for an import-triggered downloader. Our results come from two complementary analyses: a) a manual search on GitHub across 2{,}113 repositories and b) a large-scale scan of 12.3M index entries using a deobfuscating AST scanner (GOAST) that we implemented. As a result, we identified 2{,}289 malicious versions of legitimate Go modules. We demonstrate that purely GitHub-centric searches fail to identify the full extent of the compromise and are only effective for as long as the affected code is present on the platform. Moreover, our proxy-based measurements of the takedown-remediation gap reveal that among artifacts later found to be GitHub-unobservable (i.e., removed or suspended), at least 99.4\% remained retrievable via Go proxy. Following our disclosure, GitHub has removed 684 malicious repositories and the Google Go team has remediated 1{,}377 module versions. 

\end{abstract}
\section{Introduction}

The Go ecosystem serves as a backbone for today's software infrastructure \cite{GoogleCloudGo,CapitalOneGo,jetbrains_go}. Projects ranging from cloud orchestration to low-level system tools often pull modules directly from GitHub \cite{go_cloud}.

However, developers operate under the dangerous assumption that public code is safe by default \cite{Activestate2022,ENISA2021}. Mature ecosystems (e.g., npm and PyPI) provide clear evidence of adversaries exploiting this trust at scale \cite{Sonatype2023,Cisa2025npm,NPM2018,PyTorch2022,Zimmermann2019,Ladisa2023}. A recent in-the-wild campaign \cite{RedditGoAttack} shows adversaries now applying established OSS abuse techniques to Go’s unique execution and distribution model.

Attackers carried out this campaign by repackaging legitimate module names under their own accounts. They embedded obfuscated downloader code into paths that run upon import, while also manipulating social signals to speed up adoption. By fragmenting and rebuilding strings, the malware successfully hides its payload chain. Once developers import the compromised package, they unknowingly trigger a silent fetch-and-execute routine. On the social side of the repository, coordinated starring boosts perceived legitimacy and opens up star-network pivots. The attackers heavily exploit this metric, even though it is a manipulable and volatile signal \cite{He2024FakeStars,Du2020}. Go’s module distribution path actively helps sustain the attacker’s persistence strategy. Built to ensure reproducibility, the public proxy keeps cached copies of modules long after they disappear from their original hosting repositories \cite{GoProxy,SocketGoProxyBackdoor}. Ultimately, simply cleaning up the hosting layer is not enough to truly remediate the ecosystem.

Measuring such campaigns is challenging if one relies only on hosting-layer visibility. This GitHub-centric view suffers from two constraints: discovery reach limits what investigators can find, and hosting-layer volatility determines what data survives until analysis. Go, however, provides a reliable distribution-layer view via \texttt{index.golang.org} \cite{GoIndex} and \texttt{proxy.golang.org} \cite{GoProxy},  as module versions leave a stable footprint upon entering this proxy-mediated distribution path. We capture the campaign’s operation and ecosystem-scale persistence through a two-stage study uniting both of these complementary observation channels. Stage 1 relies on GitHub-centric manual discovery (March--June 2025). This initial phase isolates a high-confidence set of malicious repositories ($|M|=2{,}113$). Stage 2 then expands to the distribution layer. We collected data from the \texttt{index.golang.org} feed between July 2024 and July 2025, yielding a dataset of more than 12 million entries. Because labeling these candidates demands hard evidence at the code level, we run them through GOAST (Go AST-based Scanner for Threat), a static analysis engine designed to be both lightweight and semantically aware. Ultimately, these steps expose a confirmed distribution-layer footprint of $|G|=2{,}289$ malicious module versions. 

This paper answers three questions---(RQ1) How does the campaign operate at scale in Go? (RQ2) What does a GitHub-centric view miss? (RQ3) Does hosting-layer takedown imply ecosystem remediation?---and makes the following contributions:

\begin{itemize}
    \item \texttt{Ecosystem-scale measurement of an active Go campaign.}
    Drawing on a two-stage dataset of manual discovery ($M$) and proxy/index-scale measurement ($G$), we characterize the campaign’s automation and two-wave evolution (RQ1). We also measure how the discovery reach gap and hosting-layer volatility cause a GitHub-centric view to systematically under-scope the actual footprint (RQ2).

    \item \texttt{Quantification of the takedown--remediation gap using proxy evidence.}
    We uncover the actual module availability lingering long after original repositories disappear by matching proxy retrieval logs against GitHub snapshots (RQ3). This shows hosting-layer disappearance cannot guarantee distribution-layer unavailability.

    \item \texttt{Responsible disclosure and remediation outcomes.}
    Prior to publication, we proactively disclosed all identified threats to the relevant platform maintainers. This resulted in GitHub taking down 684 live malicious repositories and the Go team remediating 1{,}377 module versions from their proxy.
\end{itemize}

\section{Background}
The proliferation of open-source software has made supply chain security an increasingly important field of research. Although attacks targeting ecosystems such as npm and PyPI are relatively well established \cite{Sonatype2023,Cisa2025npm,NPM2018,PyTorch2022,Zimmermann2019,Ladisa2023}, recent campaign examples show that attackers are altering these approaches to take advantage of structural or architectural features of other environments \cite{GitHub2025Maven,Ohm2020,Cesarano2024GoSurf}. As a result, the Go ecosystem has become one specific target for a new, more stealthy class of supply chain attacks that can evade existing defenses and remain available through the distribution infrastructure of the ecosystem.

\subsection{Go Module Distribution, Reproducibility, and Persistence}
The Go ecosystem provides strong reproducibility guarantees via \textit{proxy.golang.org} \cite{GoProxy} and a global checksum database \textit{sum.golang.org} \cite{GoChecksum}. These methods are highly reliable for developers but also leave an additional problem: once a module version of a malicious variant is published and cached, its immutability, which guarantees reproducibility, can make it persist over time. 

This is particularly important since if an attacker releases a malicious version before it is found, then it can be cached and continue to be served even when there is no longer any hosting repository in place (i.e., removed or suspended) \cite{SocketGoProxyBackdoor}. Thus, the mere removal of a GitHub repository does not prevent it from propagating throughout the distribution pipeline, creating a critical ``\textit{takedown $\neq$ remediation}'' gap in Go's ecosystem.

Importantly, what we can observe at Go’s distribution layer is shaped by the public proxy. Not every repository on GitHub appears in the module index (\texttt{index.golang.org}). Rather, the module versions that become visible through  \textit{proxy.golang.org} are mirrored by the index \cite{GoProxy,GoModuleMirror}, which means that a repository can exist at the hosting layer without appearing in the index at all. 

As our later analysis shows, campaign footprint and persistence can be fundamentally understated with reliance on GitHub visibility alone. For this reason, understanding malicious modules in Go therefore requires proxy/index-scale visibility beyond repository-level signals alone.

\begin{figure}[h!]
    \centering
    \includegraphics[width=\linewidth]{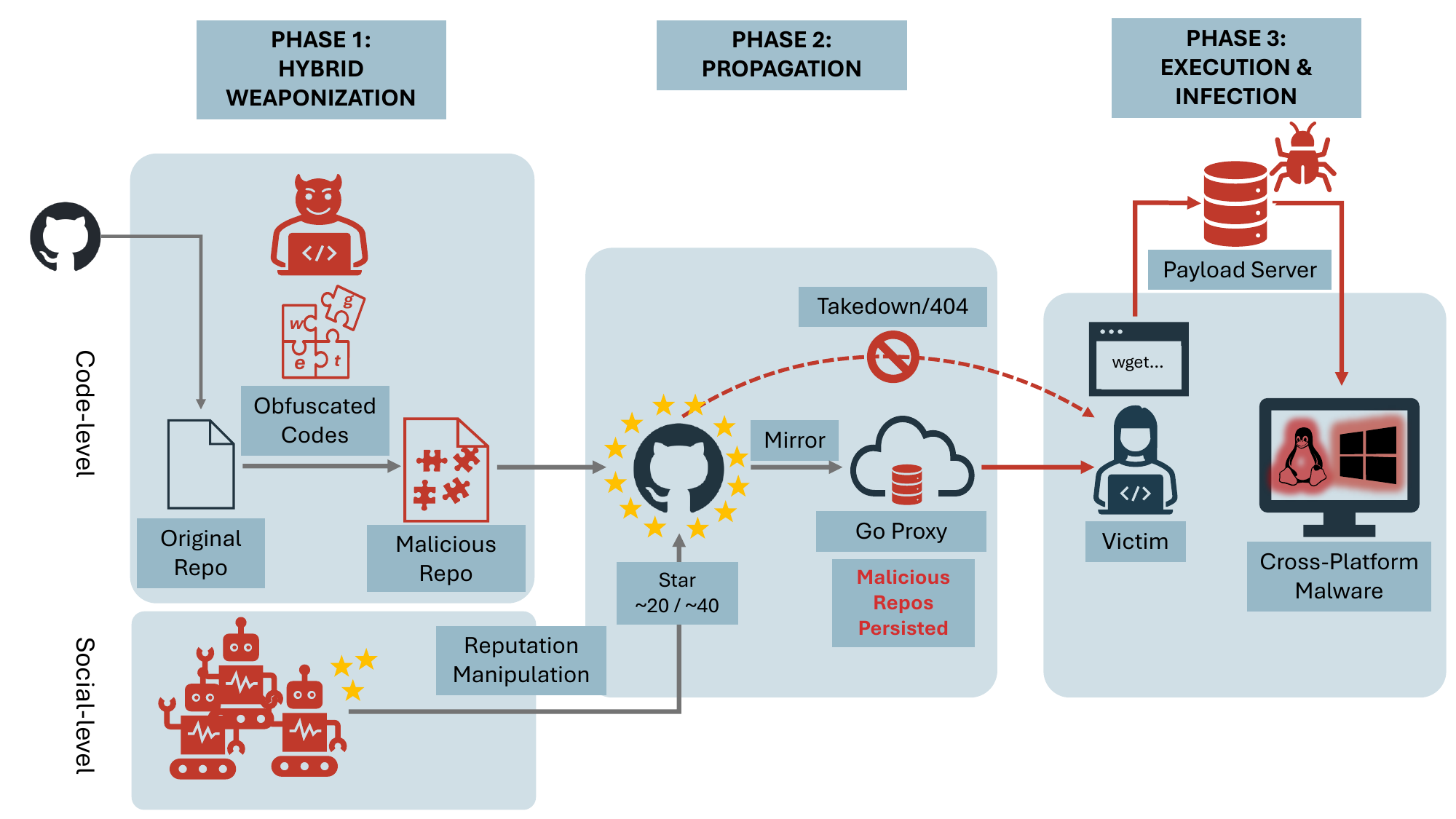}
    \caption{End-to-end campaign roadmap in the Go ecosystem.}
    \label{Fig_overview_final}
\end{figure}

\subsection{A Hybrid Threat: Adapting OSS Attack Techniques to Go}
Fig.~\ref{Fig_overview_final} illustrates onboarding techniques and import-time initializers in Go, with the downstream persistence stage following from the proxy’s immutability (\S2.1). Well-known attack techniques from open source software ecosystems have been adapted to Go by combining code obfuscation methods such as string fragmentation with social manipulation techniques such as GitHub star inflation \cite{He2024FakeStars,Du2020,Ohm2020}.

Unlike ecosystems that rely on explicit installation-time hooks, Go’s execution semantics create a subtler attack vector. In Go, package-level variable initializers execute implicitly and in declaration order when a package is imported \cite{GoImport}, and thus, any side effects in those areas automatically execute, making malicious behavior difficult to distinguish from benign logic. 

By exploiting the global variable initialization vector, identified by previous work \cite{Cesarano2024GoSurf}, attackers combine code-level obfuscation with social-level manipulation to increase campaign reach while complicating rapid triage. Crucially, the intention of the obfuscation may not need to involve perfect concealment: once surfaced during investigation, suspicious structure and intent can often be recognized rather quickly, and GenAI/LLM-aided code interpretation can lead to faster triaging.

In practice, even a brief hosting window can be enough for adversaries to onboard victims and seed modules into normal development pipelines. In the meantime, artificial star inflation may speed up onboarding of promoted subset and platform visibility remains as a weak marker of ecosystem-wide prevalence. After all, once a malicious version has been cached by Go’s distribution infrastructure, its impact can outlive the original hosting repository: temporary exposure at the hosting level may become persistent downstream availability through proxy-cached, immutable module versions \cite{GoProxy,SocketGoProxyBackdoor}.

\subsection{Limitations of Platform Signals and Hosting-Layer Volatility}
This challenge is further complicated by the constraints of hosting-layer signals and policies. At scale, platform enforcement and reporting pipelines can lag behind automated abuse \cite{Darkreading2024,GitPolicy,Apiiro2024MaliciousRepos,ZeroFox2025},
and social signals such as GitHub stars are manipulable \cite{He2024FakeStars,Du2020}.
Moreover, malicious infrastructure on hosting platforms is ephemeral: repositories tend to be suspended or deleted, adding hosting-layer volatility and hindering retrospective GitHub-centric measurement.

Instead, to reduce these observation biases, we supplement manual discovery with proxy/index-scale measurement. As we demonstrate later, this distribution-layer perspective hinges on neither platform popularity nor ongoing repository accessibility, allowing us to measure a campaign footprint stable at the distribution layer even when hosting-layer evidence vanishes.

\subsection{Scope and Responsible Disclosure}
Our analysis is preliminary: We do not execute untrusted code and do not include repository metadata in the core detection logic. We collect hosting-platform metadata only for post hoc, snapshot-based visibility measurements, not for detection. We found and archived payload binaries in order to reproduce them, but their behavioral analysis is out of scope. For dataset construction and large scale analysis, we rely on \texttt{index.golang.org} for a time-ordered feed of module versions as they are visible to the public Go module proxy, instead of relying on API-limited repository crawling to access public ecosystem at large.

We also disclosed our findings responsibly to the Google Go team and GitHub. As a result, GitHub removed 684 live malicious repositories, while the Google Go team remediated 1{,}377 malicious module versions.

\section{Methodology}

\subsection{Overview: Two-stage Discovery and Measurement}
We use a two-stage workflow to expand manual campaign discovery into proxy/index-scale measurement (M → G). To build a reliably verified baseline ($M$, where $|M|=2{,}113$), Stage 1 works outward from known public reports to map the malware's structural habits. Stage 2 of our study translates those exact habits into the core detection logic for GOAST. With GOAST scanning the artifacts retrieved through Go proxy (\texttt{proxy.golang.org}) from candidates surfaced by Go index (\texttt{index.golang.org}), we gauge the distribution-layer footprint of the campaign and collect dataset $G$ ($|G|=2{,}289$). As a strict rule, hosting metadata only informs our visibility analysis (RQ2 and RQ3). We completely ignore platform-level signals when identifying the malware itself.

\subsection{Stage 1: Manual Discovery ($M$) and Pattern Extraction}
Stage 1 manually gathers dataset $M$ ($|M|=2{,}113$) and captures the structural signatures of the attack. Capturing these techniques is essential, as they dictate how GOAST searches the proxy in Stage 2. A community discussion regarding malicious Go modules masquerading as legitimate repositories sparked our initial search \cite{RedditGoAttack}. Taking clues from that report, we queried GitHub to find the attackers' early footprint. We specifically queried the platform for execution-related fragments, relying on hits for \texttt{exec.Command}, \texttt{/bin/sh}, \texttt{-c}, \texttt{.Start()}.

\subsubsection{Stargazer-based Expansion.} 
Every seed repository shared a distinct footprint of artificial engagement. The attacker-controlled profiles boosting our initial targets routinely starred other malicious modules. This shared activity gave us a natural pivot point. We checked the broader star histories of these accounts and uncovered previously hidden repositories.
We then subjected each new discovery to a strict verification: a keyword-based pre-filter isolated .go files jointly containing \texttt{exec.Command}, a shell-invocation token, \texttt{.Start()}, and a \texttt{[]string declaration}, after which we reconstructed each flagged call's obfuscated arguments and labeled it malicious only when the recovered command matched the fetch-and-execute pattern---wget or curl retrieving and immediately executing a payload over HTTP(S). This per-candidate check was automated to keep manual effort low and directly informed the detection rules later formalized in GOAST (\S3.3). Iterating this process between March and June 2025, and across different time windows corresponding to distinct campaign waves, initially surfaced 2{,}124 repositories. Social connections drove this discovery phase; we did not expand the set by searching for structural code clones. As discussed in \S3.4, later validation excluded 11 malformed injection attempts, and all subsequent analyses therefore use a cleaned manual set $M$ of 2{,}113 repositories. The manual cost here lies in seed selection and social-graph pivoting, which fundamentally bounds scalability and motivates the ecosystem-wide automated search of Stage 2 (\S3.3).

\subsubsection{Patterns Extracted from Manual Investigation.} 
\begin{figure}[t] 
    \centering
    \includegraphics[width=\linewidth]{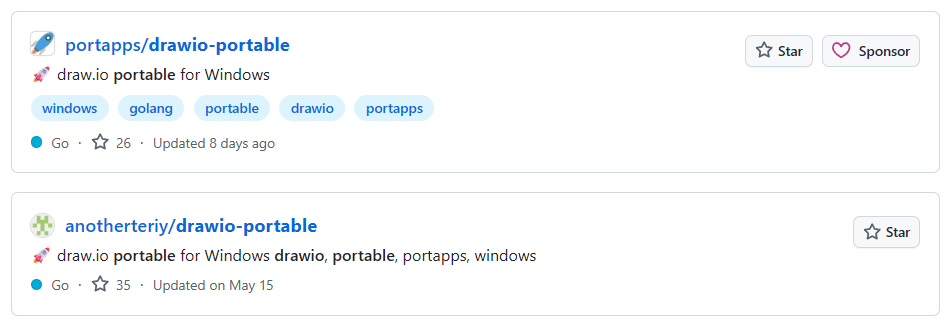}
    \caption{Visual comparison between the legitimate drawio-portable repository (top) and its malicious clone (bottom).}
    \label{fig:clone}
\end{figure}

The samples in dataset $M$ share a distinct runtime behavior: they rebuild hidden network commands from fragmented string slices. Once the malicious code concatenates these strings, it triggers exec.Command to launch the payload. It achieves this by calling /bin/sh -c on Linux environments and cmd /C on Windows. Once reconstructed, the commands reveal a predictable fetch-and-execute routine: downloading a payload via wget or curl and immediately running it. This specific behavior becomes the foundational detection rule for GOAST (\S3.3). At the repository/social layers, we observe near-clone repackaging behavior (e.g., systematic identity replacement; Fig. \ref{fig:clone}) and coordinated starring that enables stargazer-based pivots, but these signals are used only for manual discovery and are not used by GOAST. We characterize campaign-wide automation signals at scale in \S4.1.

\subsubsection{Limitations of Manual Discovery.} While pivoting through stargazers guarantees a highly accurate starting point, it creates a skewed perspective. The method naturally favors malicious infrastructure that gains visible social traction on GitHub. Furthermore, we rapidly lose this data altogether the moment the platform suspends or deletes the offending repositories. It also does not scale to ecosystem-wide measurement, motivating Stage 2 proxy/index-scale analysis.

\subsection{Stage 2: Automated Distribution-layer Measurement ($G$)}
Stage 2 scales the manual discovery process (\S3.2) to the Go distribution layer (proxy/index scale) by automating pattern detection. This stage produces the measured set $G$ ($|G|=2{,}289$), which we later compare against $M$ to characterize the campaign's true scale and visibility structure. We structured our measurement around a clear two-step pipeline. First, we use our Stage 1 findings to pull a highly targeted list of suspicious candidates straight from \texttt{index.golang.org}. Next, we run automated static analysis to label these modules based on their actual code. We built GOAST strictly as a tool for measurement. This paper centers entirely on the resulting ecosystem-scale footprint and its broader security implications, not the engineering behind the scanner itself. GOAST is deliberately specialized rather than general: it does not aim to detect arbitrary malicious Go code, but to measure this campaign's footprint at distribution scale. Unlike syntactic linters that flag any exec.Command/Start call---too broad for prevalence measurement---GOAST reconstructs the obfuscated argument and requires a confirmed fetch-and-execute command, trading generality for the precision such measurement demands.

\subsubsection{Dataset Construction.} Mapping the campaign's true reach across the ecosystem hinges on securing the precise appearance timestamps for every malicious module. We found this essential timeline by pulling a chronological feed directly from \texttt{index.golang.org}, which serves as the backbone for our Stage 2 measurements. Turning to this public registry allowed us to circumvent the heavy restrictions of the GitHub API while ensuring our data remains fully reproducible for other researchers. Our collection efforts, spanning from July 2024 through mid-July 2025, ultimately processed a volume of 12{,}304{,}230 raw records.

The proxy index logs even small module modifications, once released as new module versions, as separate entries. We cleaned the raw feed by saving only the earliest instance of a \texttt{(name, owner, first\_seen)} tuple. This deduplication shrank the dataset down to 573{,}113 unique items. From there, we narrowed the focus using a primary tactic identified back in Stage 1. The attackers actively clone legitimate module names and re-upload them using different owner accounts, such as mirroring \texttt{portapps/drawio-portable} under \texttt{anotherteriy/drawio-portable} (Fig.~\ref{fig:clone}). Keeping only the records that show this name-sharing behavior focuses the measurement on the campaign's repackaging footprint.

Finally, to focus on the campaign timeframe, we restrict candidates to those first seen on or after March 1, 2025, yielding \textbf{60{,}232 suspicious candidates} for further inspection. For each candidate, we retrieve the corresponding module source snapshot and scan it statically with GOAST. We initially attempt all downloads through the proxy. Should that fail, we pull from the hosting repository and log the fallback. Targeting the distribution layer this way means we never have to trust hosting-platform signals alone to figure out how far the campaign actually spread.

\subsubsection{Static analysis with GOAST.} 
\begin{figure*}[t]
    \centering
    \includegraphics[width=\linewidth]{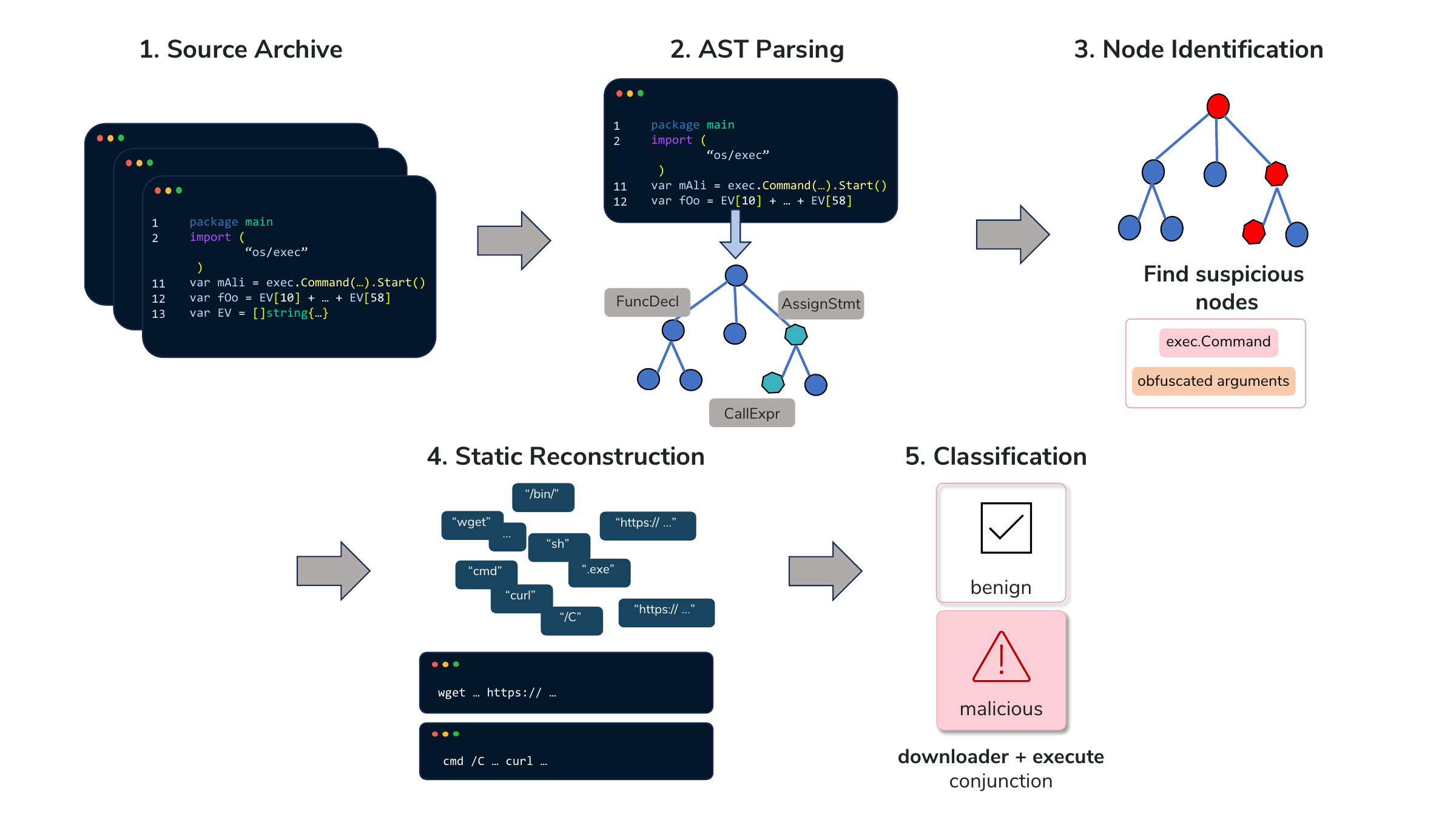}
    \caption{GOAST's five-stage detection pipeline, from source archive to a download-and-execute conjunction check (Stage 4–5). Obfuscated arguments are reconstructed before classification.}
    \label{fig:goast-pipeline}
\end{figure*}

GOAST converts Go source files into ASTs to locate process-spawning call sites, specifically targeting exec.Command combined with Start, Run, or Output. Attackers frequently hide their arguments using string concatenations or slice/index patterns. The scanner counters this by applying lightweight intra-procedural value propagation to resolve and rebuild the obfuscated inputs. To avoid flagging benign administrative scripts, detection requires a strict conjunction of signals. We label a candidate malicious only if an execution call occurs right alongside a reconstructed download instruction, such as wget or curl targeting an HTTP(S) URL. Every confirmed match records the file path, line number, and assembled snippet for auditing. As Fig.~\ref{fig:goast-pipeline} summarizes, the scanner proceeds from AST parsing through suspicious-node identification and argument reconstruction to a final download-and-execute conjunction check. \ref{app:GOAST} outlines this reconstruction concept.
 
\subsection{Validation and Dataset Relation}
Comparing the manually verified repositories ($M$) with the broader proxy footprint ($G$) requires a trustworthy baseline. We first validate GOAST's detections and then clarify how the manually verified set $M$ relates to the broader proxy footprint $G$.

\subsubsection{Validation.} We confirmed that GOAST successfully identifies every archived repository in the cleaned manual set $M$ through purely static analysis. During validation, we revisited the 2{,}124 repositories initially surfaced in Stage 1 and found that 11 corresponded to malformed injection attempts whose malicious Go files fail to parse correctly (e.g., misplaced import declarations), causing GOAST to skip them. We therefore treat these as failed attack attempts rather than valid malicious artifacts and exclude them from the final manual dataset, yielding $|M|=2{,}113$. To refine the 60{,}232 Stage 2 candidates, we ran an automated scan that flagged 2{,}297 items, eventually arriving at 2{,}289 confirmed cases after manually weeding out eight false positives. All eight are legitimate installation routines that fetch and run trusted, first-party tools (e.g., the official Docker, Caddy, SDKMAN, and Homebrew scripts) using the same curl/wget-into-shell construction as the malicious payloads. Since GOAST is content-based and ignores host reputation by design (\S3.1), it cannot separate a trusted source from a hostile one; we cleared these by confirming first-party download targets in explicit, user-invoked command paths rather than import-time initializers. This is the expected failure mode of a detector specialized to one campaign's signature, and excluding them yields |G| = 2{,}289 at 99.65\% precision. 

\subsubsection{Relationship Between $M$ and $G$.} Although $M$ and $G$ are collected through different observation channels, they capture the same campaign. Their overlap is $O=M \cap G$, with $|O|=390$. We computed this overlap by matching Stage 1 repositories and Stage 2 artifacts on the repository owner and name. A perfect one-to-one match between $M$ and $G$ is not feasible by design. Dataset $M$ grows by following social connections on the hosting platform. Dataset $G$, on the other hand, strictly logs modules that actually reached the proxy infrastructure. Because of this fundamental difference, proxy artifacts frequently lack GitHub social ties entirely or survive long after their original repository disappears. We quantify the resulting visibility gaps in \S4.2.

\section{Findings}
In this section, we unpack the campaign across both phases of our investigation. We start with the manually verified dataset ($M$) before expanding to the broader proxy-level measurement ($G$). 
Unless noted otherwise, all counts below refer to unique malicious module versions in $G$ (\S 3.3). Notably, we use metadata from the hosting platform strictly to evaluate ecosystem visibility; it plays no role in our actual malware detection process.

\subsection{RQ1: Campaign Characterization and Automation Signals}

From our Stage 1 manual investigation, a clear tactical pattern emerges. The campaign relies on cloning legitimate module names and republishing them under attacker-controlled accounts. To ensure the malware runs without suspicion, the attackers embed their payloads within Go's native initialization routines, triggering execution the moment a victim simply imports the package. Our Stage 2 measurement clarifies that the behaviors we observe in manual triage are actually present across the larger ecosystem. Extending our focus down to the entire set of proxy-retrieved artifacts ($G$) results in a strong consistency in the campaign's payloads and operational signals.

\subsubsection{Two Prominent Waves and Evolving Obfuscation/Delivery.}
Through manual analysis, we uncovered two key campaign waves, separated clearly by index timestamps and the different forms the reconstructed commands took. The attackers altered their obfuscation style and payload delivery, yet preserved the same core execution behaviors.

\paragraph{Wave 1 (March 2025):}
To build the actual Linux commands, the injected malware pieces together specific fragments from a \texttt{[]string} array. This index-driven concatenation hides the raw instructions, and upon assembly, the code triggers a classic fetch-and-execute operation, pulling a remote script and piping it straight into the shell.

\paragraph{Wave 2 (May 2025):}
Attackers expanded delivery to include Windows branches and diversified obfuscation by combining fragmented literal concatenation with the index-based reconstruction observed in Wave 1. The Linux routines stick to piping fetched scripts directly to the shell. The Windows logic, conversely, composes \texttt{cmd /C} sequences specifically designed to drop and run PE payloads.

\begin{table}[t]
\centering
\footnotesize
\caption{Summary of two-wave evolution.}
\label{tab:wave_compare}
\begin{tabular}{p{0.24\linewidth} p{0.35\linewidth} p{0.35\linewidth}}
\toprule
\textbf{Characteristic} & \textbf{Wave 1} & \textbf{Wave 2} \\
\midrule
Time window & Mar 2025  & May 2025  \\
Targeting platform  & Linux &  + Windows \\
Execution & \texttt{/bin/sh -c} &  + \texttt{cmd /C} \\
Obfuscation style & Slice indexing & + Literal concatenation \\
\bottomrule
\end{tabular}
\end{table}

\subsubsection{Import-time execution remains the invariant.}
Even as the campaign evolved across two waves, the attackers never abandoned their core execution method: spawning processes using \texttt{exec.Command(...).Start()}. Instead of exposing the commands in cleartext, the malware dynamically pieces them together from scattered fragments right before execution. The adversaries ensure the malware runs silently by hooking into code paths that execute automatically during the import phase. By hiding their logic inside global variable initializers and similar structures, they effectively camouflage the malicious side effects as standard, benign initialization logic.

\subsubsection{Campaign-wide automation signals: rigid conformity beyond code.}
While the artifacts flagged by GOAST are identified purely by static code evidence, ecosystem-scale analysis of the flagged artifacts provides independent signals consistent with a single automated pipeline.

\paragraph{Owner naming and size distributions.}
Owner names are rigidly formatted: 84.7\% ($1{,}939/2{,}289$) contain 11--13 lowercase characters. Artifact sizes are also subject to strict constraints: 95.8\% ($2{,}194/2{,}289$) do not exceed 500\,KB, and no artifact exceeds 2\,MB.
Importantly, these features are not used by GOAST for detection but are independent proof that the flagged artifacts are the result of well-structured, automation-driven campaign.

\paragraph{Delivery infrastructure and payload rotation.}
The deobfuscated commands from dataset $G$ resolve to an unexpectedly narrow infrastructure footprint. In total, the campaign relies on just 20 distinct domains (\ref{app:domains}). The payload links themselves also share a rigid blueprint. Every download request is formatted as \texttt{https://<domain>/storage/<token>/<token>}.
We repeatedly extracted this exact URL pattern from module versions that had no obvious connection to each other. The Linux code initiates a standard fetch-and-pipe via wget. The accompanying Windows logic, however, constructs curl commands specifically designed to download and run executable files.
We verified this centralized infrastructure by downloading the actual payloads from these URLs and analyzing their SHA-256 hashes. This tracking revealed two entirely different maintenance strategies. The Linux endpoints served a static ELF stager, returning the exact same file hash during every retrieval. The Windows endpoints, however, actively rotated their malware. Through daily polling, we mapped a shifting sequence of unique Windows PE binaries that repeated on a strict 31-day cycle (\ref{app:indexedPEs}).

Taken together, the narrow domain pool and rotating payload schedule point to a single, centrally managed, and automated delivery pipeline operating at scale.

\subsubsection{Takeaway (RQ1).} The attackers mass-produce repackaged modules across the ecosystem under a strict obfuscation formula. While this tactic easily defeats naive scans, its rigid structure enables reliable static reconstruction. At this scale, manual triage quickly becomes inadequate, motivating proxy/index-scale measurement for visibility and remediation analysis (RQ2--RQ3).

\subsection{RQ2: What Does a GitHub-Centric View Miss?}
Stage 2 was initially motivated by a prevalence question: we generalized the code-level patterns learned from manual investigation and measured how widely the campaign had spread across the Go distribution layer. Comparing the GitHub-centric manual view ($M$) with the proxy/index-scale view ($G$) shows two distinct blind spots that explain why hosting-only analysis is systematically incomplete. \textbf{Miss \#1} captures artifacts in the proxy/index footprint that a visibility-driven GitHub workflow fails to discover (discovery reach gap), whereas \textbf{Miss \#2} captures artifacts that remain in the proxy/index footprint but cannot be inspected on GitHub at snapshot time because their repositories have already been removed or suspended (hosting-layer volatility). We quantify both misses below.

\subsubsection{Miss \#1: Discovery Reach Gap (manual vs.\ proxy/index).}
As discussed in \S3.4, although the two sets ($|M|=2{,}113$ and $|G|=2{,}289$) have similar magnitude, their overlap $O$ is limited: $|O|=|M \cap G| = 390$ (18.4\% of $M$ and 17.0\% of $G$), leaving $|M \setminus G|=1{,}723$ and $|G \setminus M| = 1{,}899$. Thus, within the proxy/index-measured footprint, our GitHub-centric manual workflow did not capture 1{,}899 out of 2{,}289 (83.0\%) artifacts.

This non-overlap does not indicate a ``recall failure'' of either method. Our methodology starts with GitHub social pivots in Stage 1 before expanding to evaluate what the proxy network actually distributes in Stage 2 (\S2.1). Consequently, the difference $M \setminus G$ represents GitHub repositories that never made it into the broader proxy ecosystem. In contrast, $G \setminus M$ highlights the opposite: modules actively circulating in the proxy that our manual pivot strategy failed to find.

\subsubsection{Miss \#2: Hosting-layer Volatility (GitHub Unobservable).}
Working backward from the distribution-layer footprint reveals a major timing gap. By the time we captured our GitHub snapshots, attackers or the platform had already deleted many of the source repositories, making the corresponding artifacts impossible to inspect on the hosting layer. We label an artifact as \emph{GitHub-observable} if its repository is accessible at snapshot time; otherwise, it is \emph{GitHub-unobservable}.

During the June snapshot for dataset $M$, we successfully fetched metadata for 1{,}515 repositories. As a result, out of the 2{,}113 total artifacts, 598 (28.3\%) were already GitHub-unobservable category. By the August snapshot for dataset $G$, we retrieved metadata for just 703 repositories, leaving 1{,}586 of the 2{,}289 artifacts (69.3\%) in our proxy-measured footprint already GitHub-unobservable. Practically speaking, any investigation relying strictly on GitHub at that time would be unable to inspect the repositories backing these modules.

\begin{table}[h]
\centering
\footnotesize
\caption{Proxy/index footprint $G$ (August snapshot) by manual discovery reach (rows; Miss \#1) and GitHub visibility (columns; Miss \#2). Percentages are relative to $|G|=2{,}289$.}
\label{tab:rq2_misses}
\begin{tabular}{l r r r}
\toprule
\textbf{Measured Set ($G$)} & \multicolumn{2}{c}{\textbf{GitHub Visibility (August Snapshot)}} & \\
 \cmidrule(lr){2-3}
 & \textbf{Observable} & \textbf{Unobservable} & \textbf{Total} \\
\midrule
In $M$ ($M \cap G$) & (a) 276 (12.1\%) & (b) 114 (5.0\%) & 390 (17.0\%) \\
Not in $M$ ($G \setminus M$) & (c) 427 (18.7\%) & (d) 1{,}472 (64.3\%) & 1{,}899 (83.0\%) \\\midrule
\textbf{Total} & 703 (30.7\%) & 1{,}586 (69.3\%) & 2{,}289 (100\%) \\
\bottomrule
\end{tabular}
\end{table}

\subsubsection{Intersecting the Two Blind Spots.}
Table \ref{tab:rq2_misses} breaks down the distribution of these unobserved artifacts. The data heavily skews toward a complete lack of visibility. We found that 1{,}472 artifacts occupy the ``double-miss'' space (d), making up 64.3\% of the overall footprint. These modules evaded the manual search ($G \setminus M$) while also losing their hosting-layer presence before the August snapshot. Moreover, GitHub-unobservability is concentrated in the portion missed by manual discovery. The GitHub-unobservable rate is 29.2\% in $M \cap G$ (b: 114/390), whereas it jumps to 77.5\% for artifacts outside manual coverage (d: 1{,}472/1{,}899).

Importantly, hosting-layer volatility is distinct from discovery reach. It is not only that some artifacts are missed by visibility-driven pivots (Miss \#1); the hosting-layer view itself can disappear over time (Miss \#2). Among the 703 artifacts still observable at the August snapshot (Table \ref{tab:rq2_misses}), manual discovery failed to identify a large portion of the malware (c: 427/703), highlighting the discovery reach gap (Miss \#1). At the same time, hosting-layer volatility continues even within a fixed set. Within the overlap subset $O=M \cap G$, GitHub-unobservability increased from 74 (19.0\%) in the June snapshot to 114 (29.2\%) by the August snapshot, showing that Miss \#2 operates independently of Miss \#1. Together, these results demonstrate that a GitHub-centric view is limited in two distinct ways and motivate RQ3, where we measure the remediation surface that remains at the distribution layer after hosting-layer disappearance.

\subsubsection{Takeaway (RQ2).}
A purely GitHub-centric approach simply cannot capture the full extent of the threat. This limited view naturally misses much of the proxy/index footprint (Miss \#1). Furthermore, by the time of analysis, a large fraction of that footprint is no longer visible on GitHub due to hosting-layer volatility (Miss \#2). These two blind spots drive the core investigation of RQ3. If a repository gets deleted from the hosting layer, is the threat truly remediated from the Go ecosystem?

\subsection{RQ3: Measuring the Takedown--Remediation Gap in the Wild}
\begin{figure}[t]
  \centering
  \includegraphics[width=\linewidth]{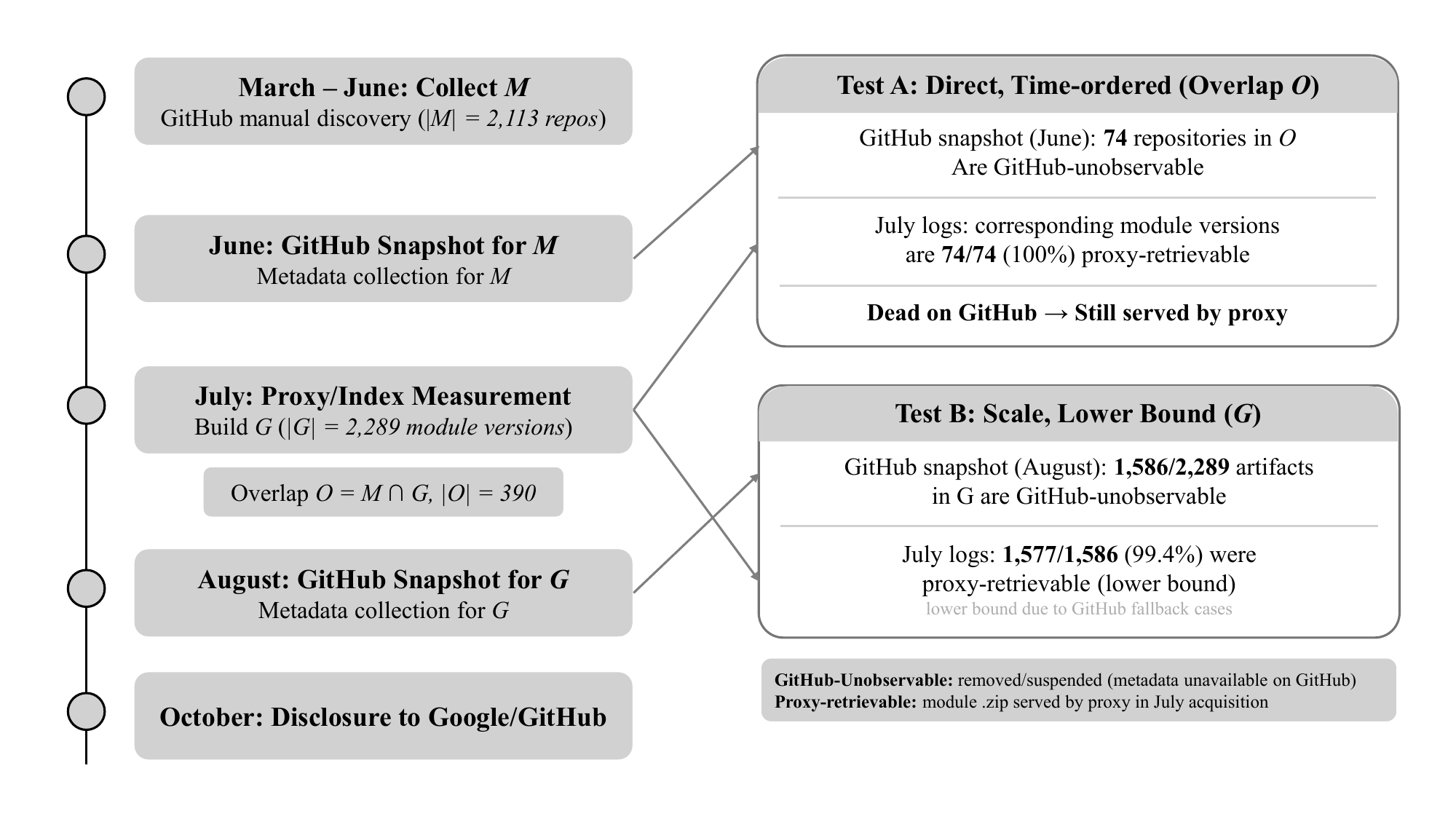}
  \caption{Timeline and two complementary tests used in RQ3 to quantify the takedown–remediation gap. Test A uses the overlap $O = M \cap G$; Test B scales to $G$.}
  \label{fig:rq3_timeline}
\end{figure}

In RQ3, we investigate the relationship between hosting-layer takedowns and actual ecosystem remediation. Specifically, once a malicious repository becomes GitHub-unobservable, we test whether its corresponding module versions still survive within the proxy-mediated distribution path. Fig. \ref{fig:rq3_timeline} summarizes our measurement points and two complementary tests.

\paragraph{Setup.}
Here, as in RQ2 (\S4.2), \emph{GitHub-unobservable} refers to repositories no longer visible on GitHub, while \emph{proxy-retrievable} refers to module versions whose \texttt{.zip} files were served by \texttt{proxy.golang.org} in our July acquisition logs.

\subsubsection{Evidence A (Overlap $O$): Proxy After GitHub Takedown.} The intersection $O=M \cap G$ ($|O|=390$) provides a natural time-ordered view of the ecosystem. Dataset $M$ logs the baseline GitHub visibility in June, while $G$ records the proxy retrieval outcomes a month later. Within this shared group, 74 repositories had already vanished from the hosting platform during the June snapshot. Yet, despite being confirmed dead on GitHub, every single one of their corresponding module versions---74 out of 74 (100\%)---remained fully accessible via \texttt{proxy.golang.org} in July. Because GitHub unobservability was confirmed before the proxy retrieval attempt, this provides direct, time-ordered evidence that hosting-layer takedown does not necessarily eliminate distribution-layer availability.

\subsubsection{Evidence B (Across $G$): Scale of Proxy Accessibility.} We next quantify how broadly this gap applies across the full proxy/index-measured footprint $G$. As reported in RQ2 (\S4.2), 1{,}586 of 2{,}289 artifacts (69.3\%) were GitHub-unobservable at the August metadata snapshot. Cross-referencing this set against our July acquisition logs, we find that 1{,}577 of these 1{,}586 artifacts (99.4\%) were proxy-retrievable in July. The remaining 9 artifacts were retrieved via GitHub fallback during acquisition. Since our pipeline is proxy-first with GitHub fallback, these 9 cannot be confirmed as proxy-retrievable from the logs. Therefore, 1{,}577 should be interpreted as a lower bound on the number of artifacts that both (i) later became GitHub-unobservable and (ii) had already entered the proxy-mediated distribution path during our measurement window. \footnote{We did not systematically re-query \texttt{proxy.golang.org} after the August GitHub snapshot; thus, this result characterizes proxy accessibility at our July measurement point.}

\subsubsection{Takeaway (RQ3).}
In this campaign, hosting-layer takedown did not imply ecosystem remediation. Our analysis confirms that proxy-accessible modules frequently persist long after their GitHub sources disappear (Evidence A). We measure this lingering threat at scale, exposing the remediation gap across the Go ecosystem (Evidence B). Operationally, this means that large-scale repackaging campaigns can rely on short-lived GitHub infrastructure to seed persistent distribution-layer availability, making GitHub-only response workflows systematically incomplete.

\section{Discussion}
\subsection{Implications for Incident Response and Measurement} Combining RQ1–RQ3 shows that brief hosting-layer exposure can be sufficient for malicious module versions to enter the proxy/index distribution path; after caching, subsequent hosting-layer volatility mainly removes upstream context while leaving a residual distribution-layer remediation surface. Therefore, incident response should enumerate affected module versions using proxy/index evidence and treat GitHub metadata as auxiliary when repositories remain observable. Looking at either M or G in isolation misses the full picture, since each channel reflects a different phase of the attack.

\subsection{Design Implications}
These visibility gaps point to two concrete directions. One is a queryable, version-level source-liveness signal (e.g., reachable/unreachable with a last-checked time) to support rapid triage when upstream repositories disappear. Common public hosts could support such signals, though freshness, coverage, and cost trade-offs remain practical challenges. Another is a machine-readable remediation inventory at the module-version level. Go already provides an OSV-based vulnerability database \cite{GoVulnDB} for identifying known vulnerable modules, but remediation tracking remains a distinct operational need. Recording the path, version, and timing of each distribution-layer remediation action would provide an audit trail that supports footprint scoping even when GitHub repositories are no longer observable.

\subsection{Limitations and Future Measurement}
For our Stage 2 measurement, we rely exclusively on Go index (\texttt{index.golang.org}) entries captured within our specific collection timeframe. Any module versions entering the proxy or index outside this window may not appear in G. As a result, our reported footprint serves as a temporal snapshot rather than an exhaustive campaign inventory. Because Go index reflects versions that became visible through the public proxy, activity routed through private proxies or direct VCS fetches may also fall outside our distribution-layer view. In addition, our proxy accessibility evidence in RQ3 is tied to our July acquisition logs. Evidence B is therefore a conservative lower bound, since artifacts retrieved via GitHub fallback cannot be confirmed as proxy-retrievable from these logs, and we did not conduct systematic post-snapshot re-queries of Go proxy (\texttt{proxy.golang.org}). While our proxy and index metrics confirm distribution-path availability, the true scale of downstream victim impact remains unquantified. Since this study focuses strictly on the distribution layer, future research should analyze downstream dependency graphs at scale to measure actual developer adoption and exposure.

\subsection{Ethical Considerations}
As described earlier, all active malicious repositories identified in this study were responsibly disclosed to the Google Go team and GitHub. To prevent any unintentional propagation from this paper and to protect other researchers, all URLs and repository names presented in this work have been truncated or normalized to a non-functional form. We did not execute any malicious code or interact with victim systems. All malicious artifacts were handled in an isolated environment for static analysis only. No personal data was collected or processed during artifact collection. To support reproducibility while limiting the risk of misuse, GOAST's detection logic and the de-identified module-version metadata underlying our measurements are available from the authors upon reasonable request.

\section{Related Work}
\paragraph{Ecosystem-scale supply chain attacks and measurement.}
When examining large open-source package ecosystems, prior work shows that their deeply interconnected structure can amplify security risks \cite{Zimmermann2019}. Complementary work systematizes recurring attack vectors into broad taxonomies \cite{Ladisa2023}, while registry-scale studies capture the broader prevalence and structural weak points that case-by-case analyses often miss \cite{Duan2021MalOSS}. These works converge on a shared insight: tracking supply-chain exposure and attacker behavior demands ecosystem-scale measurement instead of relying purely on surfaced incidents. Building on this line of work, we apply this measurement logic to Go, where the observable footprint is proxy/index-mediated rather than purely repository-visible.

\paragraph{Campaign automation and Go-specific execution semantics.}
Investigating these threats also requires understanding how attackers operate at scale. Manual reviews of malicious packages uncover clear structural patterns. By analyzing these datasets, researchers found that attackers routinely rely on obfuscation and form interconnected campaign clusters through code reuse or shared patterns \cite{Ohm2020}. However, when targeting Go specifically, these campaigns must also adapt to the language's execution model. The GoSurf study categorizes these unique threats, drawing attention to the danger of \texttt{init()} functions and global initializers that run automatically upon import \cite{Cesarano2024GoSurf}. We observe this exact dynamic in our study (RQ1). The attackers behind the campaign actively merge code obfuscation and rapid repackaging with Go's native import mechanics to ensure their payloads execute silently.

\paragraph{Hosting-layer volatility and proxy/index visibility.}
Measuring the true scale of these campaigns ultimately comes down to finding a reliable vantage point. Attackers routinely manipulate signals at the hosting layer, so treating those platforms as a secure foundation is risky. As recent work demonstrates, adversaries frequently boost short-lived malware using coordinated fake stars \cite{He2024FakeStars}. When these repositories inevitably vanish, tracking their past footprint on GitHub becomes almost impossible. These observations motivate our analysis of the limits of hosting-layer visibility (RQ2). The Go distribution layer naturally counters these hosting-layer gaps, as documented in recent defense research. In real-world incidents, malicious pseudo-versions can remain retrievable from the proxy even after upstream cleanups \cite{Cesarano2025GoLeash}. Recent studies analyze these proxy and index logs to map out vulnerability lifecycles and patch delays across Go \cite{Hu2024GoVulnLifecycle}. Leveraging these proxy signals, we measure the campaign's observed footprint and quantify the resulting takedown-remediation gap (RQ3).

\section{Conclusion}
In this work, we presented an ecosystem-scale measurement of an automation-driven supply-chain campaign targeting the Go ecosystem. We paired GitHub-centric manual discovery ($|M|=2{,}113$ repositories) directly with proxy/index-scale measurement ($|G|=2{,}289$ module versions). As our data shows, limited discovery reach and hosting-layer volatility force a GitHub-only view to systematically under-scope the footprint.

Crucially, our proxy evidence exposes a practical takedown-remediation gap in the wild. We found that module versions stayed retrievable through the proxy-mediated distribution path long after their repositories became GitHub-unobservable. We responsibly disclosed our findings prior to publication, leading to cross-layer remediation (684 repositories removed and 1{,}377 module versions remediated). These results highlight that durable remediation in Go requires coordination across hosting and distribution layers.

%\begin{credits}
%\subsubsection{\ackname} A bold run-in heading in small font size at the end of the paper is
%used for general acknowledgments, for example: This study was funded
%by X (grant number Y).

%\end{credits}

\clearpage
\appendix
\renewcommand\thesection{Appendix~\Alph{section}}
\renewcommand{\theHsection}{appendix.\Alph{section}}

\section{GOAST Pseudocode}
\label{app:GOAST}

\begin{algorithm}[H]
\scriptsize
\SetAlgoLined
\caption{GOAST: AST-based Two-Pass Detection with Context Accumulation}
\KwIn{Go source repository $R$}
\KwOut{MALICIOUS or CLEAN, Alert set $\mathcal{A}$}

\textcolor{blue}{\tcp{ Phase 1: Parse AST and Build Global Context}}
$\mathcal{T} \leftarrow \{ \text{ParseAST}(f) \mid f \in R \}$, $\mathcal{C}_g \leftarrow \text{ExtractGlobalContext}(\mathcal{T})$\;

\textcolor{blue}{\tcp{Phase 2: Context-Aware AST Traversal (global and function scopes)}}
$\mathcal{A} \leftarrow \emptyset$, $\mathcal{Q} \leftarrow \emptyset$, $hasExec \leftarrow \text{false}$, $hasObf \leftarrow \text{false}$\;

\ForEach{FuncDecl $F$ in $\mathcal{T}$}{
    $\mathcal{C}_l \leftarrow \emptyset$\;
    \ForEach{AST node $n$ in \textbf{traverse}($F$)}{
        \Switch{$n$.type}{
            \Case{AssignStmt}{
                $val \leftarrow \text{EvalAST}(n.rhs, \mathcal{C}_g, \mathcal{C}_l)$, $\mathcal{C}_l[n.lhs] \leftarrow val$\;
                \lIf{IsObfuscated($val$)}{$\mathcal{A} \leftarrow \mathcal{A} \cup \{(F, n.line, val)\}$, $hasObf \leftarrow \text{true}$}
            }
            \Case{CallExpr where IsExecCommand($n$)}{
                $cmd \leftarrow \text{ReconstructArgs}(n.args, \mathcal{C}_g, \mathcal{C}_l)$\;
                \lIf{$cmd \neq \bot \land$ IsSuspicious($cmd$)}{$\mathcal{A} \leftarrow \mathcal{A} \cup \{(F, n.line, cmd)\}$, $hasExec, hasObf \leftarrow \text{true}$}
                \lElse{$\mathcal{Q} \leftarrow \mathcal{Q} \cup \{(n, \mathcal{C}_l)\}$}
            }
        }
    }
}

\textcolor{blue}{\tcp{Pass 2: Resolve Deferred Calls}}
\ForEach{$(n, \mathcal{C}_{saved}) \in \mathcal{Q}$}{
    $cmd \leftarrow \text{ReconstructArgs}(n.args, \mathcal{C}_g, \mathcal{C}_{saved})$\;
    \lIf{$cmd \neq \bot \land$ IsSuspicious($cmd$)}{$\mathcal{A} \leftarrow \mathcal{A} \cup \{(n.file, n.line, cmd)\}$, $hasExec, hasObf \leftarrow \text{true}$}
}

\Return $(hasExec \land hasObf)$ ? MALICIOUS : CLEAN, $\mathcal{A}$\;

\BlankLine
\SetKwProg{Fn}{Function}{:}{}
\Fn{\text{EvalAST}$(n, \mathcal{C}_g, \mathcal{C}_l)$}{
    \lIf{$n$.type is BasicLit or Ident}{\Return ResolveValue($n$, $\mathcal{C}_g$, $\mathcal{C}_l$)}
    \lElseIf{$n$.type is BinaryExpr}{\Return EvalAST($n.X$) $\oplus$ EvalAST($n.Y$)}
    \lElseIf{$n$.type is IndexExpr}{\Return EvalAST($n.X$)[$n.Index$]}
    \lElseIf{$n$.type is CompositeLit}{\Return ExtractSlice($n$)}
    \lElse{\Return $\bot$}
}

\Fn{\text{IsObfuscated}$(value)$}{
    \Return length($value$) $\geq$ 40 $\land$ (shortStrings($value$) / length($value$)) $> 0.75$\;
}

\Fn{\text{IsSuspicious}$(cmd)$}{
    \Return RegexMatch($cmd$, \texttt{(curl|wget).*https?://})\;
}

\label{alg:goast}
\end{algorithm}

\clearpage

\section{Observed (neutralized) domains}
\label{app:domains}

The following malicious domains were observed across multiple repositories. 
They are presented in neutralized form to prevent accidental access.

\begin{Verbatim} [fontsize=\scriptsize, frame=single]
(1) monsoletter[.]icu
(2) kavarecent[.]icu
(3) kaiaflow[.]icu
(4) hyperwordstatus[.]icu
(5) uniscomputer[.]icu
(6) infinityhel[.]icu
(7) nexofleer[.]icu
(8) mantrabowery[.]icu
(9) liquitydevve[.]online
(10) carvecomi[.]fun
(11) kaspamirror[.]icu
(12) numerlink[.]online
(13) metalomni[.]space
(14) alturastreet[.]icu
(15) requestbone[.]fun
(16) vanartest[.]website
(17) nymclassic[.]tech
(18) steemapi[.]site
(19) sharegolem[.]com
(20) 185[.]100[.]157[.]127
\end{Verbatim}

\noindent
All domains are intentionally neutralized 
(e.g., replacing “.” with “[.]”) to prevent accidental access or redistribution.

\section{Date-indexed PEs} 
\label{app:indexedPEs}
\begin{Verbatim} [fontsize=\scriptsize, frame=single]
(1) 172635b6912b06dc13f7473fc94664e9cf9dd814d7c69e7c33131d10a6ac3905
(2) b533f73e87e7f59d4c3f8a01c46f336eacede421cf53338fccf84760aba65e9c
(3) e645e9153e098ae99b35a2030ac4bcdf9d3baecc88580f933d569727173906da
(4) bef9bd50bc7817ff59478f7d23bf20fd487ae68e6b507d1b04c46210e500dffa 
(5) 42f3f9d2684328575847f3115fcd6f759cc47b0f21b3d4fea480de0f34a1e947
(6) 46541bf1a29fa238e26c74cd7a0085cc85ce2949494a8c7ba676f03cb439fb7c
(7) 92aee7ea946a89c7134e8d35a178dddedf8fb54da653f54d46d595c0b70fb362
(8) 6e8d8ff146d1b31f49bdc3f863b7515224fc6fed657ff05280e3f52b8bbc5eef
(9) b7dda0fccec2f0927bfbccb1cbfa77a2238b8e7788c327cc8915150e71413325
(10) 76435cce067c5c43d94dfdae46b212fa89e979f66e6facbc753606dd6874259b
(11) 9e29bf7906c93cda44c978feec1469a99605d144cdaf45ae359aa5479fa2c058
(12) a258bc022d4168677aa91e2344c8a561783a99e135e9219ec34a9da43e4e1d12
(13) 58bd6ccbc8fc9da4b17b9e12ca7edbbae9fcadacd43dbd5e64485e7dc1f6a3e8
(14) bbab11ae96846126726223a408590a1e79a9d6f9c470707fd5c906cda237ed4a
(15) ba6d6571b7ab3a531f3dc1e314115c7eb27cd8bc04836fbb0bb70156ee9d15c4
(16) 66b0a8d44605742536c27387285211df72e0cc4f89bf6a5e2f7a4c8acbf2c675
(17) 755cfc893c53d3cc7fcb3859ed51efc412887331847edb1aa719dd0ffd9c14d5
(18) 732cae627b80008259cff661ebc6ef176b04bd6cc296377ce1e744e117a0726c
(19) 04771fda5a0323b730a2054ccdc55f2fb9c9cbd9380ed27e9cd2a09b58f4c59f
(20) 83a191173d7e5e5030043d59ca9b46952eed3f14e7ba64c9f0c6f752cb37adf9
(21) 09e9fb549379c9232fbe97433f3264f196b039d9b3ee53cee9c6f6c6498def56
(22) f3c9ec3823c2b612e19e6bdf6af2fdd474e35d2ff59ba94de467dd0ea81e433a
(23) 4122f6c63efbb64a88a0521c94165a655da463bb12b9fc39e5f628ccb74d56d5
(24) 69b2bc171803decdb7090d0dee9466ad175600d41776ef686e3a19a2bbcaed23
(25) 3ce29c3ae9bdf71aea1aff87cb7c87c60e2b28812cf4762d57084db618954a5d
(26) c3d7af5488699bf5bc73437dbc991668146d43c86e35e1eb5f5b89d03694867d
(27) 48c4dfa17e9055366798ddd10f1ac2983e020119334f20cbe01140c02eeb8f20
(28) 0d3d4b1fcc3f6cb59bcc413a32a6e0c425bfd429960a99854833ae999c5a6ebe
(29) 4a8bf419424ff42b736a51472d35a2c172e4c60b762c519b0b2f9eb04690726c
(30) 2e72a1bd71bf60c9b9703ced3994b060f1175e471885d5f23a5e92869d5eee51
(31) e1810ade69f505fd008e09598b6530d5238fc3c55ee5b203f1d022a6fa0cb9df
\end{Verbatim}

\end{document}